\documentstyle[11pt]{article}
\newcommand{\blankline}{\vskip .3cm}
\newcommand{\f}{\begin{equation}}
\newcommand{\ff}{\end{equation}}

\begin{document}
\centerline{\LARGE   Towards a background independent approach to
$\cal M$ theory}
 \blankline
\rm
\centerline{Lee Smolin${}^*$}
\blankline
\centerline{\it  Center for Gravitational Physics and Geometry}
\centerline{\it Department of Physics}
\centerline {\it The Pennsylvania State University}
\centerline{\it University Park, PA, USA 16802}
\vfill
\centerline{August 11, 1998}
\vfill
\centerline{ABSTRACT}
Work in progress is described which aims to construct a background 
independent formulation of M theory by extending results about background 
independent states and observables from quantum general relativity and
supergravity to string theory.  A list of principles for such a theory is
proposed which is drawn from results of both string theory and non-perturbative 
approaches to quantum gravity.  Progress is reported on a
background independent membrane field theory and on a 
realization of the holographic principle based on finite surfaces.

\vfill
\blankline
${}^*$ smolin@phys.psu.edu
\eject

\section{Introduction}

Up till this day, the different approaches to quantum gravity can be
divided into two groups, which are the background dependent and background
independent approaches.  The background dependent approaches are those in 
which the definitions of the states, operators and inner product of the theory
require the specification of a classical metric geometry.  The quantum
theory then describes quanta moving on this background.  The theory may
allow the description of quanta fluctuating around a large class of backgrounds,
but nevertheless, some classical background must be specified before any
physical situation can be described or any calculation can be done.
All weak coupling perturbative approaches are background dependent, as are
a number of non-perturbative developments.  In particular, up to this point,
all successful formulations of string theory are background dependent.

The background independent approaches are those in which no classical metric 
appears in the definition of the states, operators and inner product of the 
theory.
A classical spacetime geometry can only appear in such a formulation in
an appropriately defined continuum or classical limit.  Background independent
approaches include loop quantum gravity, dynamical triangulations
and non-commutative geometry.

One quick way to describe the present state of research in quantum gravity
is that the biggest problem 
faced by the background dependent approaches is
in getting rid of the background, while the biggest problem faced by
the background independent approaches is restoring the background through
the discovery of a good classical limit. In spite of this situation,
it is still true that almost all work in quantum gravity nowadays
is carried out strictly within the context of 
one research program or another.  One is working 
on strings, or loop quantum gravity or non-commutative geometry or 
twister theory, or perhaps something else.  The main message I want to 
convey in this essay is that this is counterproductive, and that progress from
this point on will be faster if more people can think in terms of a 
single,``quantum theory of 
gravity under construction", which will have elements of more than one of 
these programs.  If we do this we will force ourselves to discover and
bridge the link between the background dependent and 
background independent approaches.

There is a good reason to believe that this is the moment to attempt to
bridge this gap.  This is that both string theory and the
background independent approaches to quantum gravity have produced results 
which,  by 
their robustness, generality and simplicity, may be considered predictions 
about the real world.   At the same time, there is evidence 
within each of these research programs  that it is not itself the whole 
story.   This, together with the striking fact that in several cases results 
of more than one research program point to the same conclusion, suggests 
strongly that the complete theory must involve elements of both string theory 
and the background independent approaches\footnote{It should
be stressed that by background independent I mean something much
stronger than results that show that the theory is well defined as an
expansion around any background, or any background which solves
some equation.  In truly background independent formulations the metric and 
connection enter the theory only as operators, and no classical metric 
appears in the definition of the state space, dynamics or gauge symmetries.}.

One examples of this is the fact that a large number of results in string 
theory point to the existence of a new theory, called $\cal M$ theory, 
which unifies all the perturbative string theories such that all of those 
theories turn out to describe expansions around different vacua or phases 
of it\cite{duff-M,witten-M}.  Almost by 
definition, that theory requires a description which is 
background independent, as its different vacua or phases involve different 
background manifolds.  Even more striking, there are results such as 
mirror manifolds that indicate that different manifolds may be equivalent 
in this theory\cite{mirror}, or that there are processes that allow 
transitions between 
different manifolds\cite{andy-bh}.  

A theory that encompasses these phenomena cannot be based on an 
expansion around a manifold, and must therefore be background 
independent.  A key problem in string theory then must be to construct a 
background independent formulation of $\cal M$ theory.  

It is then striking that there are very few ideas for how to approach the 
construction of a background independent string or $\cal M$ theory.   
String field 
theory 
is a natural possibility but, so far, there is no completely satisfactory 
background independent formulation of closed string field theory. 
 Matrix models\cite{matrix} are so far (but see \cite{mematrix}) not only 
background dependent but restricted to light cone gauge.  
The wonderful new 
AdS/CFT correspondence\cite{juan-ads,witten-ads,other-ads}, 
for all its attractive features, also seems 
dependent on the choice of a fixed manifold.   There is an interesting 
proposal of Horava\cite{horava}, for a diffeomorphism invariant version
of $\cal M$ theory based on an $11$ dimensional Chern-Simons theory.
That theory has a large number of local degrees of freedom, and even at 
the classical level the sorting out of gauge and physical degrees of 
freedom is a difficult job that has only just been started\cite{highercs}.

It may seem that the construction of a background independent string theory
should be very difficult, given that all the versions of string theory so far
formulated are background dependent\footnote{There are of course 
non-perturbative results in string theory, but these are, so far, background
dependent.}.
However, the problem may seem difficult because one
is limiting the perspective to 
methods developed in string theory itself.   If one is willing to bring into 
string theory structures and methods discovered in  research programs 
whose very purpose has been to develop background independent methods 
for quantum theories of gravity, the problem may be easier than 
anticipated.  Chief among these has been the program to quantize 
diffeomorphism invariant theories of gravity, commonly known as ``loop 
quantum gravity"\cite{lp1,lp2,aa,spain,sn1,vol1,carlo-reviews}
\footnote{The extension of loop quantum gravity to supergravity is
described in a number of 
papers\cite{ted-sugr,sloops,othersugra}.}.

Coming to that program, it is helpful to divide the results which have been 
achieved into two sets which we may call ``kinematic" and ``dynamic".  In 
the first class are all those results that require only that we are studying 
a gauge theory, based on a connection valued in some algebra or 
superalgebra, $A$.  Such a theory is defined on 
a spatial manifold $\Sigma$ of some dimension $d$, 
in which the gauge symmetries include both ordinary gauge transformations 
valued in $A$ and diffeomorphisms of $\Sigma$.  
These include general relativity and supergravity, 
coupled to arbitrary matter fields, plus a large number of other theories, 
both topological and with local degrees of freedom.  Here the results include a complete characterization of the space of gauge invariant 
states, which turn out to  have an elegant description in 
terms of combinatorics and representation theory\cite{sn1,vol1}.

Related to the characterization of the gauge invariant states has been the 
development of techniques to construct large classes of gauge invariant 
operators\cite{spain,vol1}.  
Because of the diffeomorphism invariance, when these 
constructions succeed they produce finite operators\cite{spain}.  
Amongst those so 
constructed are, in general relativity and supergravity, operators that 
correspond to the areas of surfaces and the volumes of regions 
they bound\cite{spain,sn1,vol1,volume2}, 
as well as dynamical operators such as 
the hamiltonian constraint\cite{lp1,lp2,ham1,thomas} and the 
hamiltonian in fixed gauges\cite{me-dieter}.

This whole class of results has also been found to be consequences of a  
general and rigorous formulation of diffeomorphism invariant quantum 
field theory.  Results such as the characterization of the gauge invariant 
states in terms of diffeomorphism classes of spin networks and the 
discreteness of area and volume have thus been elevated from results of 
calculations\cite{spain,sn1,vol1,volume2} to rigorous mathematical theorems 
that depend  on rather general 
assumptions\cite{reiner,chrisabhay,gangof5,thomas}.  
These express only the fact that the
theory is based on a connection valued in a Lie algebra $A$, that the
gauge invariance includes diffeomorphisms and $A$ valued gauge
transformations and that the frame fields are constructed from the
canonical momenta of the connection (which is the case for a large
class of gravitational theories, including supergravity.)    Beyond these, 
no assumption is made about the Planck scale dynamics of the theory.

In contrast to these kinematical results, the  dynamical results are those 
that concern a particular theory, such as general relativity, and so depend 
on the form of its hamiltonian or hamiltonian constraint.  It is very 
striking in light of recent results in string theory that here there are two 
very different kinds of results, distinguished by whether or not the
cosmological constant, $\Lambda$,
vanishes.  In the case of $\Lambda \neq 0$ there is a 
completely holographic\cite{thooft,lenny-holo} 
formulation of at least one sector of 
the theory\cite{linking,hologr},  which seems to have a good 
classical limit\cite{kodama,chopinlee}. 
This is based on the construction of a set of 
physical states in a spacetime with a timelike boundary on which 
certain boundary 
conditions have been imposed.   The Bekenstein bound\cite{bek} 
is satisfied explicitly
as the dimension of the physical state space is finite and grows like
the exponential of the area of the boundary.
Moreover, the  classical limit is precisely deSitter 
or AntideSitter spacetime\cite{kodama,chopinlee}. It is thus likely relevant
to the $AdS/CFT$ correspondence, as I shall discuss in section 4.  

One interesting result of this work is that the cosmological constant
induces a quantum deformation of the gauge theory, so that the
states and operators must be described in terms of the representation
theory of the quantum group $SU(2)_q$\cite{linking,qdef}.  This makes
the case of $\Lambda \neq 0$ very different from the theory with vanishing
cosmological constant.  In the $\Lambda =0$ case, 
an infinite dimensional space  
of exact solutions to the hamiltonian constraint is
known explicitly\cite{lp1,lp2,thomas}, but the solutions have a very 
different character than for
the case $\Lambda \neq 0$.    In contrast to that case, it appears to 
be the case
that, for reasons argued in \cite{trouble}, which are reinforced by 
recent results\cite{moretrouble}, the theory very likely does not have 
a classical 
limit in which massless particles propagate.   This is true both of the
Euclidian theory in the form studied in \cite{lp1,ham1} and in the 
Lorentzian theory
in the form studied by Thiemann\cite{thomas}.  At present no way 
is known out of these
difficulties and, while there is no theorem, it appears likely that 
at least for
$\Lambda =0$, quantum general relativity is a theory that exists, and 
describes a world
with a dynamical discrete quantum geometry, but lacks a good classical limit.

Largely because of this result, for the last two years effort in this area 
has gone into attempts to modify the theory to arrive at a theory which 
does have a good classical limit.  Most of this work has involved 
formulating the 
theory in terms of sums over histories, rather than in a strictly canonical 
language.  There are in fact two distinct approaches to the path integral
for spin network states, which are intrinsically 
Euclidian\cite{mike,carlomike,louis1,johnlouis,foam,carloint,kf} 
and Lorentzian
\cite{fotini1,fotinilee,tubes,pqtubes,moretubes,renataamb,sameer,rbsg}.
Some of this has also involved extending the degrees of 
freedom, particularly by the addition of 
supersymmetry\cite{susy-progress}.

I believe that if this program is to succeed, it must  become the same 
as the search for  a background 
independent form of $\cal M$ theory.    The argument for this is very 
simple.  Whatever quantum gravity is, if it succeeds it must have a 
classical limit.  If it is to reproduce ordinary quantum field theory it  
must also have a sensible expansion around the classical limit, which is 
by definition a perturbative theory of quantum gravity.  The only good 
perturbative quantum theories of gravity that we know of are perturbative 
string theories.  In fact, there seem to be a large number of these, but 
what is important is that no successful perturbative quantum theory of 
gravity has ever been found that was not a string theory.  

There are in fact good reasons to believe that any successful perturbative 
theory of quantum gravity must involve extended objects, whose high 
energy behavior is that of string theory.  This is because any quantum 
theory of gravity must be finite, which means that there is a fixed length 
scale, $l_{Pl}$ which marks the transition between phenomena well 
described by classical general relativity and those described by quantum 
gravity.  However, the theory we want must also have a classical limit 
corresponding to Minkowksi spacetime.  The perturbation theory around 
that limit must have Poincare invariance.

This brings us face to face with a problem, which is in whose frame is
$l_{Pl}$ to mark the boundary between the classical and quantum 
description of geometry?  It seems that there can be phenomena which are on 
one side of the line for me that are at the same time well on the other 
side for you, if our 
relative velocity is close enough to $c$.  Thus, it seems that there is a 
contradiction between the requirement that our theory be Poincare 
invariant around one classical limit and that the theory has a physical 
length scale that marks the boundary of the classical domain.

We encountered this problem in loop quantum 
gravity\cite{un}, trying to extend the 
results on the existence of gravitons in the long 
wavelength limit\cite{gravitons,jc} 
(in the frame of one observer) to a Lorentz covariant result.   On the other 
hand, it is resolved in string theory, and in a very beautiful way 
discovered by Thorne\cite{thorne} and 
Klebanov and Susskind\cite{ks,lenny-holo}.   Their 
arguments also constitute part of the evidence that string theory, which 
is based initially on the assumption that spacetime is continuous, actually 
points to a discrete picture of quantum geometry.   Their work  shows 
that the only way the apparent paradox can be resolved is if the 
excitations of the gravitational field are extended objects, which scale in 
energy as strings do.  I believe that this constitutes a very strong 
argument that any quantum theory of gravity that succeeds
will have weakly coupled excitations that behave as strings, even
if strings are not among the fundamental degrees of freedom.

Thus, if there is an extension of loop quantum gravity that has a well 
defined classical limit, it must have a regime which is described in terms of 
a perturbative string theory.  Ergo, it must {\it be} a background independent 
formulation of string theory. This argument holds for
{\it any} approach to quantum gravity,  including non-commutative 
geometry.

Non-commutative geometry\cite{ncg} is a third approach to quantum 
gravity that has progressed greatly in the last decade.  It is by definition 
a background independent approach, as the basic idea is to replace the 
background manifold by algebraic generalizations of a certain set of 
diffeomorphism invariant observables, which are the spectrum of the 
Dirac operator\cite{ncg}.  Thus, the three approaches 
also share the emphasis 
on spinorial and fermionic structures, which was anticipated in much 
earlier work of Penrose\cite{roger-sn}, Finkelstein\cite{finkelstein} 
and others.

Interestingly enough, in the last year, non-commutative geometrical 
structures have turned out to be fundamental for both 
string theory\cite{ncs} 
and loop quantum gravity\cite{qg3}.
 
It is then possible that a background independent theory is at hand which is a 
synthesis of string theory, loop quantum gravity and non-commutative geometry.  
In the 
following section I will outline briefly the picture of space and time
at the Planck scale that comes form combining the results of these
different approaches.     In the 
conclusion I mention briefly some work in progress directed towards realizing the 
picture presented here.

\section{Principles for background independent $\cal M$ theory}

If we assume that the robust results of string theory and the 
background independent approaches
are all true, we arrive at a picture of quantum spacetime that may be summarized
by a small number of statements.  These may be taken to be principles that
a theory that unified these different approaches would have to satisfy.
Given what we
know presently, these are likely to characterize any successful background 
independent
quantum theory of gravity.

 \begin{enumerate}

\item{\bf The holographic principle}  The basic idea of the holographic 
principle\cite{thooft,lenny-holo} is that in 
quantum gravity states and observables should be associated only with 
boundaries of 
regions of spacetime.  This idea has actually emerged in two different 
contexts, first in 
work by Louis Crane on the relationship of topological quantum field 
theory (TQFT) to 
loop quantum gravity\cite{louis-preholo} and then in the papers of 
Œt Hooft\cite{thooft} and 
Susskind\cite{lenny-holo}.  The latter proposal has been
developed primarily in string theory, 
while the proposal of Crane has inspired several developments on the 
background independent side
\cite{linking,pluralistic,lotc,relational,fotini2,toposholo,louis1,johnlouis,foam}. 
 
There have been so far constructed three explicit 
realizations of the holographic principle.  In historical order these are,

1)  Quantum general relativity with finite boundaries and a cosmological 
constant\cite{linking,pluralistic,hologr}. 

2)  The matrix models\cite{matrix}.

3)  The $AdS/CFT$ correspondence\cite{juan-ads,witten-ads,other-ads}

These are sufficient to show that the idea, surprising as it may seem at first, is
completely realizable within the theory we are attempting to construct.  
Furthermore, given these different realizations, one way to search for
a link between the background independent and dependent aproaches
is to investigate relationships between the different versions of the
holographic hypothesis they give rise to.

\item{ \bf Quantum spatial geometry is discrete and non-commutative.}
String theory, loop quantum gravity and non-commutative geometry all point to the 
conclusion that the geometry of space is discrete at Planck scales.  These realize 
earlier speculations by many pioneers of the field such as Penrose's spin 
networks\cite{roger-sn}.    

The basic kinematic result of non-perturbative diffeomorphism invariant
quantum field theory is that, while the metric at a point cannot be well 
defined, operators can be constructed that correspond to the areas of surfaces 
and the 
volumes they contain\cite{spain}.  These are finite and diffeomorphism invariant 
and have
discrete, computable spectra in a large class of theories including general 
relativity
and supergravity, with arbitrary matter couplings
\cite{spain,sn1,vol1,volume2,gangof5}.  

The corresponding basis of diffeomorphism invariant states correspond to 
diffeomorphism classes of spin networks for the kinematical gauge group $H$,
which for gravity and supergravity is $SU(2/{\cal N})$.  These are graphs whose
edges are labeled by representations and nodes by intertwiners.  The result is a
picture of quantum geometry that is discrete, based on representation theory and
combinatorics.

Results from string theory that point to a discrete quantum geometry are 
described in 
\cite{thorne,ks,string-discrete,discrete}.   

Evidence that the discrete quantum geometry is also non-commutative
has emerged in both non-perturbative quantum 
gravity\cite{henrijose,qg3} and
string theory\cite{ncs}.  Conversely, as it makes perturbative divergences finite,
noncommutative geometry also points to the discreteness of the quantum
geometry\cite{ncg,a-per}.

\item{ \bf Excitations are extended objects.}  The basic evidence for this is that,
as just mentioned,  the only good perturbative theories of quantum gravity
we know of are string theories.  However, it is also the case that perturbations
of background independent histories in a large class of theories are
associated with $1+1$ dimensional worldsheets, that must reproduce 
perturbative string theory if the classical limit exists\cite{stringsfrom}. 

In recent years it has been understood that the extended objects of string
theory include D-branes of various dimensionalities.  These results point to
the conclusion that strings and branes may be equally elementary.
However this does not mean that in the final, background independent theory
there will not be fundamental degrees of freedom which can be identified.
These are likely to be connected with a purely quantum description of the
background geometry, whose excitations will be then strings and branes.

\item{ \bf Consistency requires supersymmetry.}  All 
good perturbative theories of quantum gravity so far constructed
are supersymmetric\footnote{For a possible 
counterexample, see \cite{kachrusilver}.}   Thus, any sensible background 
independent 
quantum theory of gravity must either incorporate supersymmetry fundamentally 
or it must have a 
mechanism whereby supersymmetry spontaneously emerges in the perturbative limit.  
At present no such mechanism is known.  
Unless one is discovered we must 
build supersymmetry into the background independent theory.

\item{ \bf Spacetime is relational.}  Observables associated with classical general
relativity with cosmological boundary conditions measure relations between physical 
fields.  Points have no intrinsic meaning and are only identified through the 
coincidence of field values.  The diffeomorphism invariance of the classical 
theory is 
thus an expression that that theory is background independent (up to the 
specification 
of the topology of the manifold.)

Any background independent form of quantum gravity must be able to reproduce 
general relativity as a classical limit, which means it must incorporate (if indeed 
extend)  diffeomorphism invariance.  This means that the interpretation of any such 
theory must be based, as is the interpretation of classical general relativity, on
relational concepts of space and 
time\cite{relational,lotc}.  It is this fundamental point that makes it 
inconceivable that the final form of $\cal M$ theory could be expressed in terms 
of any 
particular classical background.

\item{ \bf Histories have dynamical causal structure.}  A corollary of the 
last point is that the causal structure of spacetime is a dynamical variable, which evolves 
dynamically.
That this is the case in general relativity is a direct concequence of 
diffeomorphism
invariance.  This principle must then extend to any background independent 
form of 
$\cal M$ theory.

The evolution of a discrete quantum spatial geometry must then give rise to a 
discrete 
dynamical causal structure. For the case of spin network states, the 
study of the
structures that arise has been initiated recently 
by Markopoulou\cite{fotini1,fotini2}. The
extension of this structure to a form suitable to $\cal M$ theory is under
development\cite{tubes,pqtubes,toposholo}.

Once a background independent quantum theory of gravity is formulated in
terms of histories with dynamical causal structure, the question arises as to
how to study the continuum limits of such systems.  This is necessary to understand 
the existence and properties of the different classical limits of the theory.

It is clear that since the causal structure is fluctuating, the continuum 
limits cannot be 
studied in the usual context of equilibrium second order critical 
phenomena, as the 
relevance of this phenomena to quantum field theory depends on the possibility of 
making a global Euclidian rotation, common to all histories.  The question then
arises as to what kind of critical phenomena might characterize the 
continuum limit of 
theories with dynamical causal structure.

A natural conjecture, discussed in \cite{fotinilee}, is 
that the answer is non-equilibrium critical 
phenomena.  As noted in \cite{fotinilee,fotini1,tubes,leestu} directed 
percolation, studies of the growth of soap bubbles and other non-equilibrium critical phenomena offer models which may 
be interpreted as dynamical causal structures.  The important difference is that
the histories are weighed by complex amplitudes rather than probabilities. What is 
needed is then a study of critical phenomena associated with what might be called 
quantum directed percolation problems, which are directed percolation problems in 
which the weight of a history is complex.  

Another advantage of non-equilibrium critical phenomena as a paradigm for the
continuum limit of background independent quantum theory of gravity 
is that there are 
cases in which no fine tuning is required.  This ``self-organized 
criticality\cite{per}²
is a good feature for theories of quantum cosmology to have as it may resolve the
embarrassing situation in which the existence of the classical world requires fine
tuning of parameters.  Preliminary studies of the classical limit of 
theories of fluctuating causal structure through the use of
the analogy with non-equilibrium critical phenomena 
are described in \cite{leestu,sameer,rbsg}. 

\end{enumerate}

\section{Current directions}

The basic hypothesis of this essay is that the results and conjectures
we have just outlined may all describe nature, in spite of having emerged from 
different research programs.  In fact, there is no reason the results
of string theory and the background independent approaches may not be
compatible as these programs cover different domains of quantum gravitational 
phenomena.  The question is then whether these can be combined to give one 
theory of quantum gravity that describes all domains.  As I have argued, 
this must be the same as the question of constructing a background independent 
form of string, or $\cal M$ theory.

Several projects are underway, which aim towards this goal.  I mention two of 
them very briefly.   

\subsection{Background independent membrane dynamics}

The matrix models describe a background dependent form of membrane dynamics, 
in a fixed gauge\cite{matrix}.   It is unfortunately, difficult to extend 
the matrix models even to a lorentz covariant form, for reasons described 
in \cite{mematrix}.  Thus it seems unlikely they will yield a background 
independent theory.  A background independent form of membrane dynamics was 
then proposed\cite{tubes} using an extension of the background independent form 
of dynamics that has been studied recently for spin network states of quantum 
gravity\cite{fotini1}.
The theory was  applied to $(p,q)$ string networks in \cite{pqtubes} and the 
application to $\cal M$ theory is studied in \cite{moretubes}.  The theory is 
purely quantum mechanical,  and the two dimensional surfaces which may be 
considered the constant time slices of the membranes are constructed purely 
algebraically.    The theory involves first of all the choice of an algebra 
$\cal A$ whose representation theory allows the construction of a finite 
dimensional space of conformal blocks associated to every two dimensional 
manifold with genus $g$, ${\cal V}_{{\cal A},g}$.  To define a form of 
$\cal M$ theory, $\cal A$ may be taken to be a superalgebra
with $32$ fermionic charges\cite{moretubes}.  The
hilbert space of the whole theory is taken to be,
\f
{\cal H} = \sum_g {\cal V}_{{\cal A},g}
\ff
These two surfaces are background independent membranes.  As there is to begin 
with no background there are no embedding coordinates, but there are states 
associated with the representation theory of $\cal A$.  These may also be 
considered to be states of Chern-Simons theory in a three manifold in 
the interior of the two surface.  

Time evolution is defined by an operator that generates local changes
in the topology of the surface and the state.   The result may be called 
a background independent membrane field theory.  The rules given 
in \cite{tubes,moretubes}, which extend those proposed for quantum gravity 
in terms of spin networks\cite{fotini1} result in the construction of purely 
quantum histories, which however have both causal structure and 
many-fingered time.  The time slices, which are defined algebraically, are 
associated with a basis of states in $\cal H$.   Each history has an 
amplitude, which is the product of an amplitude for the local moves.  
The dynamics is given by specifying the forms of these amplitudes as 
described in \cite{tubes,moretubes}.

A key problem for such a theory is the classical limit, which is, as I've
argued above, a problem in non-equilibrium critical phenomena.  Some studies 
of this problem are underway\cite{leestu,rbsg}.    

Some general results about the perturbation theory in such a framework are known.  
In particular, it can be argued that perturbations of these abstract histories 
are given by a discrete field theory defined on a timelike two surface embedded 
in the history\cite{stringsfrom}.  When there is a classical limit, the two 
dimensional theory must contain the masslesss modes, hence it must define in 
the continuum limit a consistent perturbative string theory.

How is the physical interpretation of such a theory to be given, in the 
absence of any background?  As described in \cite{tubes} this is done in 
terms of information projected on two surfaces embedded in the two surfaces 
on which the states are defined.  Thus the theory is holographic, 
by construction. The relationship between the distribution of 
information on these surfaces and the causal structure is 
rather intricate, and may be described using a mathematical formulation 
developed in \cite{fotini2}.

\subsection{Holography on finite surfaces}

As stressed in the last section, 
in a background independent theory there will be no asymptotic 
classical region, and hence, if there is to be a holographic formulation, 
it must be defined on finite surfaces inside the universe.  One way to 
approach the construction of such a theory is to extend the holographic 
formulation of
quantum general relativity given in \cite{linking} to a candidate for a 
form of $\cal M$ theory.
This may be done by extending the algebra of observables and states 
of the boundary theory from one suitable for general relativity to one 
suitable for $N=8$ supergravity.  Such a formulation will be described 
in \cite{n=8}, based on a general form for quantum theories of gravity 
as constrained topological quantum field theories developed in \cite{hologr}.
Of course, these are not the only possible approaches to a background 
independent form of string theory.  They may for example be criticized 
in that, while supersymmetry can be easily included, it seems optional 
from the point of view of the background independent formulation.  
It is possible that the main role of supersymmetry in such formulations 
is that it guarantees the existence of the classical limit.  However, 
it is also possible that supersymmetry plays an even more fundamental 
role, which is yet to be revealed.

\section*{ACKNOWLEDGEMENTS} 

I am grateful to Arivand Asok, Roumen Borissov, Sameer Gupta, Stuart Kauffman, 
Fotini Markopoulou and Chopin Soo for collaborations in the work described.  
Conversations with John Baez,  Tom Banks, Per Bak, Shyamoli Chaudhuri, 
Louis Crane, Murat Gunyadin, Renata Kallosh,
Juan Maldecena, George Minic,  Maya Paczuski, 
Carlo Rovelli, Lenny Susskind, Ergin Szegin and Edward Witten were most 
helpful in formulating these ideas.   
This work was supported by
NSF grant PHY-9514240 to The Pennsylvania State
University and a NASA grant to the Santa Fe Institute.  In addition,
the author would like to express his thanks for support from the
Jesse Phillips foundation.

\end{document}